\documentstyle[sprocl]{article}
\input{psfig}
\def\Journal#1#2#3#4{{#1} {\bf #2}, #3 (#4)}

\def\PRL{\em Phys. Rev. Lett.}
\def\PRD{{\em Phys. Rev.} D}
\def\APJ{{\em Astrophys. J.}}
\def\APJS{{\em Astrophys. J. Suppl.}}
\def\AA{{\em Astron. Astrophys.}}

\begin{document}
%
%
\let\ov=\over
\let\lbar=\l
\let\l=\left
\let\r=\right
\def\be{\begin{equation}}
\def\ee{\end{equation}}
%
%

\title{GRAVITATIONAL WAVES FROM ISOLATED \\
NEUTRON STARS\footnote{to appear in the Proceedings of the International
Conference on Gravitational Waves: Sources and Detectors, Cascina (Pisa),
Italy --- March 19-23, 1996 (World Scientific, in press).}}

\author{E. GOURGOULHON, S. BONAZZOLA}

\address{D\'epartement d'Astrophysique Relativiste et de Cosmologie,\\
UPR 176 C.N.R.S.,\\
Observatoire de Paris, \\
F-92195 Meudon Cedex, France}

\maketitle\abstracts{
Continuous wave gravitational radiation from isolated rotating
neutron stars is discussed. The general 
waveform and orders of magnitude for the amplitude are presented
for various known pulsars.  
The specific case of gravitational radiation resulting from
the distortion induced by the stellar magnetic field is presented.
Finally some preliminary results about
the signal from the whole population of neutron stars in the Galaxy
are discussed.}

\section{Introduction}

As compact objects with significant internal velocities, neutron
stars constitute a priori valuable gravitational wave sources.
Besides catastrophic
events such as their coalescence \cite{Blanc96} or their gravitational
collapse \cite{StarP85}, these stars can be interesting sources of
continuous wave gravitational radiation, provided they deviate from 
axisymmetry. Whereas the accompanying lecture \cite{BGPise96} focuses on
mechanisms, such as spontaneous symmetry breaking, which imply
accretion from a companion, the present lecture is devoted to 
radiation from isolated rotating neutron stars. 

\section{General considerations}

\subsection{A general formula for gravitational emission by a rotating star}
\label{s:general}

When dealing with neutron stars, the classical {\em weak-field} quadrupole
formula for gravitational wave generation is not valid. 
For highly relativistic bodies, 
Ipser \cite{Ipser71} has shown that 
the leading term in the gravitational radiation field $h_{ij}$ is  
given by a formula which is structurally identical to the 
quadrupole formula for weak-field sources \cite{MisTW73}, 
the Newtonian quadrupole being simply replaced
by Thorne's quadrupole moment \cite{Thorn80} ${\cal I}_{ij}$. 
The non-axisymmetric deformation of neutron stars
being very tiny, the total Thorne's quadrupole can be linearly decomposed
into the sum of two pieces:
$ {\cal I}_{ij} = {\cal I}_{ij}^{\rm rot} + {\cal I}_{ij}^{\rm dist} $,
where ${\cal I}_{ij}^{\rm rot}$ is the quadrupole moment due to rotation
and ${\cal I}_{ij}^{\rm dist}$ is the quadrupole moment due to the
process that distorts the star, for example an internal magnetic 
field \cite{BonaG96},
anisotropic stresses from the nuclear interactions or irregularities in 
the solid crust \cite{Haens96}.
Let us make the assumption that the distorting process has a
privileged direction, i.e. that two of the three 
eigenvalues of ${\cal I}_{ij}^{\rm dist}$ are equal. Let then $\alpha$ be the
angle between the rotation axis and the principal axis of 
${\cal I}_{ij}^{\rm dist}$ which corresponds to the non-degenerate
eigenvalue. The two modes $h_+$ and $h_\times$
of the gravitational radiation field in a transverse traceless gauge
derived from the Thorne-Ipser quadrupole formula are \cite{BonaG96}
\begin{eqnarray}
   h_+ & = & h_0 \sin\alpha \Big[
	{1\ov 2} \cos\alpha\sin i\cos i \cos\Omega t 
 - \sin\alpha {1+\cos^2 i\ov 2} \cos2\Omega t \Big] \label{e:h+,gen} \\
   h_\times & = & h_0 \sin\alpha \Big[
	{1\ov 2} \cos\alpha\sin i\sin\Omega t
	- \sin\alpha \cos i \sin2\Omega t \Big] \ , \label{e:hx,gen}
\end{eqnarray}
where $i$ is the inclination angle of the ``line of sight'' with respect
to the rotation axis and
\be \label{e:h0,eps}
    h_0 = {16\pi^2 G\ov c^4} {I\, \epsilon\ov P^2\, r} \ , 
\ee
where $r$ is the distance of the star,
$P=2\pi/\Omega$ is the rotation period of the star, 
$I$ its moment of inertia with respect of the rotation axis and
$ \epsilon := - {3/2} \, {{\cal I}_{\hat z\hat z}^{\rm dist} /  I} $  is   
the {\em ellipticity} resulting from the distortion process.
 
It may be noticed that Eqs.~(\ref{e:h+,gen})-(\ref{e:h0,eps}) are
structurally equivalent to Eq.~(1) of Zimmermann \& Szedenits \cite{ZimmS79},
although this latter work is based on a different physical hypothesis
(Newtonian precessing {\em solid} star).

From formul\ae\ (\ref{e:h+,gen})-(\ref{e:hx,gen}), it appears clearly that
there is no gravitational emission if the distortion axis is aligned
with the rotation axis ($\alpha=0\ \mbox{or}\ \pi$). If both axes are
perpendicular ($\alpha=\pi/2$), the gravitational emission is monochromatic
at twice the rotation frequency. In the general case ($0<|\alpha|<\pi/2$),
it contains two frequencies: $\Omega$ and $2\Omega$. For small values
of $\alpha$ the emission at $\Omega$ is dominant.
Numerically, Eq.~(\ref{e:h0,eps}) results in the amplitude
\be \label{e:h0,num}
    h_0 = 4.21\times 10^{-24} \ \Big[ {{\rm ms}\ov P} \Big] ^2
	\Big[ {{\rm kpc}\ov r} \Big] 
	\Big[ {I\ov 10^{38} {\ \rm kg\, m}^2} \Big]
	\Big[ {\epsilon \ov 10^{-6} } \Big] .
\ee
Note that $I=10^{38} {\ \rm kg\, m}^2$ is a representative value for
the moment of inertia of a $1.4 \, M_\odot$ neutron star. 
In the following, $I$ is systematically set to this value.  
	
The values of $h_0$ resulting from Eq.~(\ref{e:h0,num}) are given in 
Table~\ref{t:h0,pulsars} for two young rapidly rotating pulsars 
(Crab and Vela), the nearby pulsar Geminga \cite{CaBMT96} 
and two millisecond pulsars: the second \footnote{the ``historical'' 
millisecond pulsar PSR 1937+21, which is the
fastest one, is not considered for it is more than twice
farther away.} fastest one, PSR 1957+20, and the nearby millisecond pulsar
PSR J0437-4715 \cite{Johns93}. At first glance, millisecond pulsars 
seem to be much more favorable candidates than
the Crab or Vela. However, in Table~\ref{t:h0,pulsars}, $\epsilon$ is in units 
of $10^{-6}$ and the very low value of the period derivative $\dot P$ of
millisecond pulsars implies 
that their ellipticity is at most $2\times 10^{-9}$, as we shall
see in \S~\ref{s:upper}. 
 
\begin{table}[t]
\caption[]{\label{t:h0,pulsars}
Gravitational wave amplitude $h_0$ on Earth as a function of the ellipticity
$\epsilon$ for five pulsars.}
\vspace{0.4cm}
\begin{center}
\begin{tabular}{|c|c|c|c|}
\hline
name & rotation period & distance & GW amplitude \\
  & $P {\ \rm [ms]}$ & $r {\ \rm [kpc]}$ & $h_0$ \\ 
\hline
Crab & 33 & 2 &  $1.9\times 10^{-27} (\epsilon / 10^{-6} )$ \\
Vela & 89 & 0.5 & $1.1\times 10^{-27}( \epsilon / 10^{-6} )$ \\
Geminga & 237 & 0.16 & $4.7\times 10^{-28}( \epsilon / 10^{-6} )$ \\
PSR B1957+20 & 1.61 & 1.5 & $1.1\times 10^{-24} ( \epsilon / 10^{-6} )$ \\
PSR J0437-4715 & 5.76 & 0.14 & $9.1\times 10^{-25}  ( \epsilon / 10^{-6} )$
 \\ \hline
\end{tabular}
\end{center}
\end{table}

\subsection{Detectability by VIRGO} \label{s:detectability}

Let us give a crude estimate of the minimum amplitude $h_0$ 
detectable by the VIRGO interferometer. Whereas the expected amplitude
is very weak, as compared with other astrophysical processes such 
coalescences or gravitational collapses, 
one can take advantage of the permanent character of the signal
to increase the signal-to-noise ratio by increasing the observing time. 
Indeed for an integration time $T$, the signal-to-noise ratio reads
\be
    {S\ov N} = {h_0\ov \tilde h(f)} \sqrt{T} \ ,
\ee
where $\tilde h(f)$ is VIRGO sensitivity (square root 
of the noise spectral density) at the frequency $f$. 
The miminum values of $h_0$ leading to $S/N=1$ when $T=3{\ \rm yr}$ are
given in Table~\ref{t:min,h0}. In view of these values, we shall take as
a basis for our discussion that {\em in order to be detectable by VIRGO, 
a rotating neutron star must produce $h_0 > 10^{-26}$}. 

\begin{table}[t]
\caption[]{\label{t:min,h0}
Minimal amplitude $h_0$ detectable ($S/N=1$) 
by VIRGO within 3 years of integration. The values of $\tilde h(f)$ have
been taken from Giazzoto et al.\cite{Giazz95} }
\vspace{0.4cm}
\begin{center}
\begin{tabular}{|c|c|c|}
\hline
frequency  & sensitivity & detectable amplitude \\
$f {\ \rm [Hz]}$ & ${\tilde h} {\ \rm [{Hz}^{-1/2}]}$ & $\min h_0$ \\
\hline
$10$ & $10^{-21}$ & $10^{-25}$ \\
$30$ & $10^{-22}$ & $10^{-26}$ \\
$100$ & $3\times 10^{-23}$ & $3\times 10^{-27}$ \\
$1000$ & $3\times 10^{-23}$ & $3\times 10^{-27}$ \\ \hline
\end{tabular}
\end{center}
\end{table}

\subsection{Upper bounds on gravitational radiation from pulsars} 
\label{s:upper}

An absolute upper bound on $h_0$ can be derived by assuming that the observed
slowing down of the pulsar (the so-called $\dot P$)
is entirely due to the energy carried away
by gravitational radiation. Let us stress that this is not a realistic 
assumption
since most of the $\dot P$ is thought to result instead from losses via
electromagnetic radiation and/or magnetospheric 
acceleration of charged particles --- at least for Crab-like pulsars. 
However, this provides an upper bound
on the ellipticity $\epsilon$ and the gravitational wave amplitude $h_0$.
The resulting values are given in Table~\ref{t:hmax}. The five first entries
in this Table correspond to the five highest values of $h_{0,\rm max}$
among the 706 pulsars of the catalog by Taylor 
et al.\cite{TayML93}$^{\!,\,}$\cite{TaMLC95}. 

\begin{table}[t]
\caption[]{\label{t:hmax}
Absolute upper bounds on the ellipticity and the GW amplitude derived from 
the measured spin-down rate of pulsars.}
\vspace{0.4cm}
\begin{center}
\begin{tabular}{|c|c|c|c|c|}
\hline
 & \multicolumn{2}{|c|}{GW frequencies}  & max.  
	& max.  \\
 name & \multicolumn{2}{|c|}{ } & ellipticity & GW amplitude \\ 
  & $f {\ \rm [Hz]}$ & $2f {\ \rm [Hz]}$ & $\epsilon_{\rm max}$ 
 & $h_{0,\rm max}$ \\ 
\hline
Vela & 11 & 22 & $1.8\times 10^{-3}$ & $1.9\times 10^{-24}$  \\
Crab & 30 & 60 & $7.5\times 10^{-4}$ & $1.4\times 10^{-24}$  \\
Geminga & 4.2 & 8.4 & $2.3\times 10^{-3}$ & $1.1\times 10^{-24}$  \\
PSR B1509-68 & 6.6 & 13.2 & $1.4\times 10^{-2}$ & $5.8\times 10^{-25}$  \\
PSR B1706-44 & 10 & 20 & $1.9\times 10^{-3}$ & $4.2\times 10^{-25}$  \\
\hline
PSR B1957+20 & 621 & 1242 & $1.6\times 10^{-9}$ & $1.7\times 10^{-27}$  \\
PSR J0437-4715 & 174 & 348 & $2.9\times 10^{-8}$ & $2.6\times 10^{-26}$  \\
\hline
\end{tabular}
\end{center}
\end{table}

New et al.\cite{NeCJT95} have recently pointed out that if the 
mean ellipticity of pulsars is taken to be of the order of 
the $\epsilon_{\rm max}$ of millisecond pulsars, i.e. 
$\epsilon \sim 10^{-9}$ (cf. Table~\ref{t:hmax}), then the Crab pulsar 
reveals to be a much worse candidate than PSR B1957+20, as it can be seen by
setting $\epsilon = 10^{-9}$ in Table~\ref{t:h0,pulsars}: 
$h_0^{\mbox{\tiny Crab}} \simeq 2\times 10^{-30}$ versus 
$h_0^{\mbox{\tiny 1957+20}}\simeq 10^{-27}$. 

\subsection{Ellipticity required for a detectable amplitude}

Given the threshold $h_0 = 10^{-26}$ 
for detectability by the VIRGO interferometer (\S~\ref{s:detectability}), 
one can consider the 
corresponding value of $\epsilon$ resulting from Eq.~(\ref{e:h0,num}) and 
compare it with the maximum value given by the pulsar slowing down 
(Table~\ref{t:hmax}). The results are presented in Table~\ref{t:eps,min}. 
From that it can be concluded that an ellipticity as small as $10^{-8}$
leads to a detectable amplitude for the nearby millisecond pulsar
PSR J0437-4715, whereas the Crab or Vela pulsar should have an ellipticity
of the order $10^{-5}$, which is about one percent of the maximum allowable
ellipticity as given by the spin-down rate (\S~\ref{s:upper}). 

\begin{table}[t]
\caption[]{\label{t:eps,min}
Minimum values of the ellipticity required to produce $h_0=10^{-26}$ on Earth.} 
\vspace{0.4cm}
\begin{center}
\begin{tabular}{|c|cl|}
\hline
 name & min $\epsilon$ &  \\
\hline
Crab & $5.3\times 10^{-6}$  & $=7\times 10^{-3} \, \epsilon_{\rm max}$ \\
Vela & $9.1\times 10^{-6}$  & $=5\times 10^{-3} \, \epsilon_{\rm max}$ \\
Geminga & $2.1\times 10^{-5}$ & $=9\times 10^{-3} \, \epsilon_{\rm max}$ \\
PSR B1957+20 & $9.1\times 10^{-9}$  & $> \epsilon_{\rm max}$ \\
PSR J0437-4715 & $1.1\times 10^{-8}$  & $=0.4 \, \epsilon_{\rm max}$ \\
\hline
\end{tabular}
\end{center}
\end{table}

\section{The specific case of magnetic field induced distortion}

In this section, we consider the specific example when the distortion 
results from the neutron star's magnetic field. 
In this case the ellipticity is expressible as 
$\epsilon = \beta {\cal M}^2/{\cal M}_0^2$, where $\cal M$ is the magnetic
dipole moment, ${\cal M}_0 = 2.6\times 10^{32} {\ \rm A\, m}^2$
and $\beta$
a dimensionless coefficient which measures the efficiency of the magnetic
structure in distorting the star. For an incompressible fluid Newtonian body 
endoved with a uniform magnetic field \cite{GalTT84}, $\beta = 1/5$. 
For a given pulsar, $\cal M$ can be inferred from the value of $P\dot P$. 

We have developed a numerical code to compute the deformation of 
magnetized neutron stars within general 
relativity \cite{BonaG96}$^{\!,\,}$\cite{BoBGN95}.   
The solutions obtained are fully relativistic and 
self-consistent, all the effects of the
electromagnetic field on the star's equilibrium (Lorentz force, spacetime
curvature generated by the electromagnetic stress-energy) being taken into
account. The magnetic field is axisymmetric and poloidal.  

\begin{figure}
\centerline{
\psfig{figure=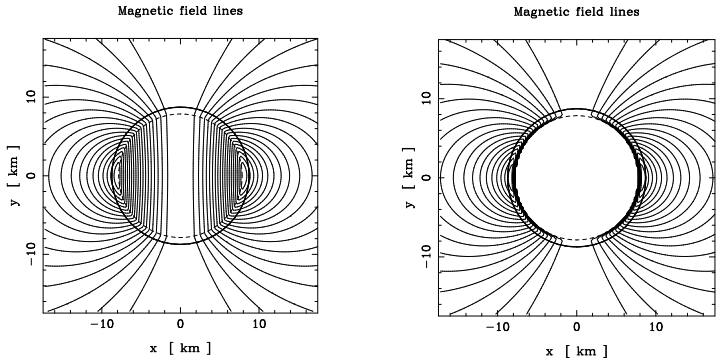,height=5cm}
}
\caption[]{\label{f:mag:conf}
Magnetic field lines generated by a current distribution localized in 
the crust of the star (left) or exterior to a 
type I superconducting core (right). The thick line denotes the star's surface
and the dashed line the internal limit of the electric current distribution
(left) and the external limit of the superconducting region (right). 
The distortion factor corresponding to these configurations is $\beta=8.84$
(left) and  $\beta=157$ (right).}
\end{figure}

The reference (non-magnetized) configuration is taken to be a $1.4\, M_\odot$
static neutron star built with
the equation of state ${\rm UV}_{14}+{\rm TNI}$ 
of Wiringa, Fiks \& Fabrocini \cite{WirFF88}. 
Various magnetic field configurations have been considered; the most 
representative of them are presented hereafter.

Let us first consider the case of a perfectly conducting interior (normal
matter, non-superconducting).  
The simplest magnetic configuration compatible \cite{BoBGN95} 
with the MHD equilibrium of the star 
results in electric currents in the
whole star with a maximum value at half the stellar radius in the equatorial plane. 
The computed
distortion factor is $\beta=1.01$, which is above the $1/5$ value of the 
uniform magnetic field/incompressible fluid Newtonian model \cite{BonaG96} 
but still very low. 
Another situation corresponds to electric currents localized in the neutron
star crust only. Figure~\ref{f:mag:conf} presents one such configuration: 
the electric current is limited to the zone $r>r_*= 0.9 \, r_{\rm eq}$. 
The resulting distortion factor is $\beta=8.84$. 

In the case of a superconducting interior, of type I (which 
means that all magnetic field has been expulsed from the superconducting 
region), the distortion factor somewhat increases.
In the configuration depicted in Fig.~\ref{f:mag:conf}, the
neutron star interior is superconducting up to $r_*=0.9\, r_{\rm eq}$. 
For $r>r_*$, the matter is assumed to be a perfect conductor carrying some
electric current. The resulting
distortion factor is $\beta=157$. 
For $r_*=0.95\, r_{\rm eq}$, $\beta$ is even higher: $\beta = 517$. 

The above values of $\beta$, of the order $10^2 - 10^3$, though much higher than
in the simple normal case, are still too low to 
lead to an amplitude detectable by the first generation of 
interferometric detectors in the case of the Crab or Vela pulsar, which 
would require \cite{BonaG96} $\beta> 10^4$.  
It is clear that the more disordered the magnetic field the higher $\beta$,
the extreme situation being reached by a stochastic magnetic
field: the total magnetic dipole moment $\cal M$ almost vanishes, in 
agreement with the observed small value of $\dot P$, whereas the mean value 
of $B^2$ throughout the star is huge. 
Note that, according to Thompson \& Duncan \cite{ThomD93}, turbulent dynamo
amplification driven by convection in the
newly-born neutron star may generate small scale magnetic fields as strong as
$3\times 10^{11}{\ \rm T}$ with low values of $B_{\rm dipole}$ outside the
star and hence a large $\beta$. 
In order to mimic such a stochastic magnetic field, we
have considered the case of counter-rotating electric currents.
The resulting distortion factor can be as high as 
$\beta=5.7\times 10^3$. 
 
If the neutron star interior forms a type II superconductor, 
the magnetic field inside the star
is organized in an array of quantized magnetic flux tubes, each tube containing
a magnetic field $B_{\rm c} \sim 10^{11} \ {\rm T}$.
As discussed by Ruderman \cite{Ruder91}, the crustal stresses induced by the pinning of 
the magnetic flux tubes is of the order $B_{\rm c} B / 2\mu_0$, 
where $B$ is the mean value of the magnetic
field in the crust ($B\sim 10^8 \ {\rm T}$ for typical pulsars). This
means that the crust is submitted to stresses $\sim 10^3$ higher than in
the uniformly distributed magnetic field 
(compare $B_{\rm c} B/2\mu_0$ with $B^2/2\mu_0$).
The magnetic distortion factor $\beta$ should increase in the same
proportion. We have not done any numerical computation to confirm this
but plan to study type II superconducting interiors in a future work.

\section{Gravitational radiation background from the whole population
of neutron stars in the Galaxy}

\subsection{The squared signal}

\begin{figure}
\centerline{
\psfig{figure=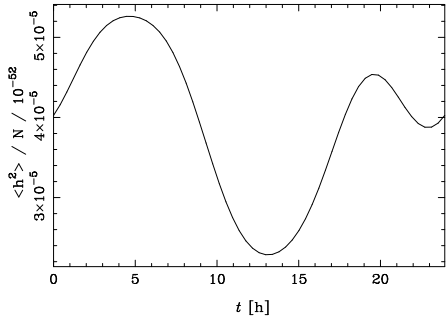,height=5cm}
}
\caption[]{\label{f:disq}   \protect\footnotesize
Total squared signal (divided by $N$) from a distribution of $N$
neutron stars concentrated in the galactic disk with a height scale of
$0.5$ kpc. The average rotation period is assumed to be 
$\bar P = 5 {\ \rm ms}$ 
and the mean ellipticity $\bar\epsilon = 10^{-8}$.} 
\end{figure}

Let $N$ be the total number of neutron stars in our Galaxy. 
The response of an interferometric detector to the 
gravitational wave field $(h_+^i,h_\times^i)$ 
[cf. Eqs.~(\ref{e:h+,gen})-(\ref{e:hx,gen})] emitted by the 
$i^{\rm th}$ neutron star is
\be
    h_i(t) = F_+^i(t)\, h_+^i(t) + F_\times^i(t)\, h_\times^i(t) \ , 
\ee
where $F_+^i(t)$ and $F_\times^i(t)$ are beam-pattern factors which 
depend on the direction of the star with respect to the detector arms. 
They vary with time because of the Earth rotation and revolution around
the Sun. This results in an amplitude 
modulation \cite{JotaD94}$^{\!,\,}$\cite{BonaG96} as well as a 
Doppler shift \cite{JotVD96}$^{\!,\,}$\cite{Grave96} of the signal.
The time-average of the total (i.e. the sum on all the galactic neutron
stars) is zero but not the time-average of the {\em squared} total signal:
\be
   \langle h^2 \rangle = {1\ov \tau} \int_{t_0}^{t_0+\tau}
	\l[ \sum_{i=1}^N h_i(t) \r] ^2 \, dt \ .
\ee
The key-point is that if the galactic neutron star distribution is not
isotropic, $\langle h^2 \rangle$ will exhibit temporal variation, with a 
period of one sidereal day. The precise shape of the signal depends
on the neutron star distribution. An example \cite{BonGG96} corresponding 
to a disk distribution is presented in Fig.~\ref{f:disq}.

It can be shown \cite{BonGG96} that the signal-to-noise ratio for
detecting this signal is
\be
    {S\ov N} = { \langle h^2  \rangle \ov 2\tilde h^2 } 
      \l( {T \ov 2 \Delta\nu } \r) ^{1/2} \ , 
\ee
where $\Delta\nu$ is the frequency bandwith and $T$ is the observation
time. It appears that such a signal might be detected if the number
of neutron stars falling in the frequency range of the VIRGO detector
is larger than $\sim 10^6 - 10^7$. 

\subsection{Number of rapidly rotating neutron stars in the Galaxy}

From the star formation rate and the outcome of supernova explosions, 
the total number of neutron stars (NS) in the Galaxy has been estimated 
\cite{TimWW96}$^{\!-\,}$\cite{Pacin96}
to be about $10^9$. The number of {\em observed} NS is
much lower than this number: $\sim 700$ NS are observed as
radio pulsars \cite{TayML93}$^{\!-\,}$\cite{TaMLC95}, $\sim 150$ as 
X-ray binaries, among which $\sim 30$ are X-ray pulsars
\cite{WhiNP95}$^{\!-\,}$\cite{vanPa95}, and a few as isolated NS, through 
their X-ray emission \cite{WalWN96}.

For our purpose, the relevant number is given by the fraction of 
these $\sim 10^9$ NS which rotates sufficiently rapidly to emit
gravitational waves in the
frequency bandwith of VIRGO-like detectors. Putting the low
frequency threshold of VIRGO to $f_{\rm min} = 5{\ \rm Hz}$ 
(optimistic value !), and
using the fact that the highest gravitational frequency of a
star which rotates at the frequency $f$ is $2f$ (cf. \S~\ref{s:general}), 
this means the rotation period of a detectable NS must be lower
than $P_{\rm max} = 0.4{\ \rm s}$. 
The NS that satisfy to this criterion can be divided in three 
classes:
(C1) young pulsars which are still rapidly rotating 
(e.g. Crab or Vela pulsars);
(C2) millisecond pulsars, which are thought to have been spun up by
accretion when being a member 
 member of a close binary system (during this phase the system 
may appear as a low-mass X-ray binary); 
(C3) NS with $P<0.4{\ \rm s}$ 
which do not exhibit the pulsar phenomenon. 

The number of millisecond pulsars in the Galaxy
is estimated \cite{Bhatt95} to be of the order $N_2 \sim 10^5$. 
The number of {\em observed} millisecond pulsars ($P<10 {\ \rm s}$)
is about 50 and is continuously increasing.   

The number of young (non-recycled) rapidly rotating NS is more difficult
to evaluate. An estimate can be obtained from the fact that the
observed non-millisecond pulsars with $P<0.4{\ \rm s}$ represent
28 \% of the number of catalogued pulsars and that the 
total number of active pulsars in the Galaxy is around $5\times 10^5$.
The number of rapidly rotating non-recycled pulsars 
obtained in this way is $N_1 \sim 1.4\times 10^5$. 

Adding $N_1$ and $N_2$ gives a number 
of $\sim 2\times 10^5$ NS belonging to the populations (C1) and (C2) 
defined above. 
This number can be considered as a lower bound for the total number
of NS with $P<0.4 {\ \rm s}$. The final figure depends on the
number of members of the population (C3). This latter number is (almost by
definition !) unknown. All that can be said is that it is lower than 
the total number of NS in the Galaxy ($\sim 10^9$).

\section*{Acknowledgments}
This work has benefited from discussions with F.~Bondu, A.~Giazotto, 
P.~Haensel, P.~Hello and S.R.~Valluri.

\end{document}